\documentclass[iop]{emulateapj}

\def\mathstacksym#1#2#3#4#5{\def#1{\mathrel{\hbox to 0pt{\lower 
    #5\hbox{#3}\hss} \raise #4\hbox{#2}}}}

\mathstacksym\lta{$<$}{$\sim$}{1.5pt}{3.5pt}
\newcommand{\FeI}{\ion{Fe}{1}}
\newcommand{\CaII}{\ion{Ca}{2}}

\newcommand{\pref}{\protect\ref}
\newcommand{\solrad}{\ifmmode{R}_{\rm S}\else${R}_{\rm S}$\fi}
\newcommand{\solmas}{\ifmmode{M}_{\rm S}\else${M}_{\rm S}$\fi}

\newcommand{\ctn}{\ifmmode\kappa\else$\kappa$\fi}

\newcommand{\term}[2]{\mbox{$\,^{#1}{\rm #2}$}}
\def\term#1 #2/{\mbox{$\,^{#1}{\rm #2}$}}
 
\newcommand{\solis}{{SOLIS}}
\newcommand{\ibis}{{IBIS}}
\newcommand{\spinor}{{SPINOR}}
\newcommand{\hinode}{{\em Hinode}}
\newcommand{\trace}{{\em TRACE}}

\newcommand{\tabone}{
\protect\begin{deluxetable}{llllll}
\tablecaption{Log of observations for 20 May 2008\label{tab:obslog}}
\tablehead{Observatory & Instrument & Observable & Center & FOV & Time
UT}
\startdata
DST & IBIS & Fe~I 630.2, Ca~II 854.2, H$\alpha$ & (-161,171) & $40\times80$&
14:29-15:05 \\
    & WL  & white light & &$40\times80$ & ditto\\
    & G band & G band & &$90\times90$ & ditto\\
\hinode{} & SP & Fe~I 630.2 fast map& (-144,210) & $107\times82$&14:30-16:39\\
 & FG & Na~I D, Ca~II H, G band & (-137,204) )& $82\times82$&14:30-16:41\\
 & EIS & 9 lines& (-126,189) & $520\times 512$& 14:30-16:37\\
  & XRT & 3 bands & (-139,169) & $130\times169$ & 14:00:16:36\\
\trace{} & & WL, 160nm, 19.5nm & (-166,199)& $512\times512$& 14:01-16:42\\
\\
DST & IBIS & Fe~I 630.2, Ca~II 854.2, H$\alpha$ & (666,-193) & $40\times80$&
15:12-15:19 \\
    & WL  & white light & & &ditto\\
    & G band  & G band & & &ditto\\
\enddata 
\tablecomments{Solar coordinates are relative to the \solis{}
 magnetogram obtained near 15:16 UT and are reliable to
 about $\pm1\arcsec$. The center of the SP and FG images reported in
 the headers before 15:05 UT were quite different from those in the table, being 
(-151,183) and (-145,176) respectively. The \trace{} image center is
 for the 19.5 nm band.  The \ibis{} commanded pointing was
 (-165,158). The pointing is approximate for the EIS instrument.
}
\end{deluxetable}
}

\newcommand{\figone}{
\begin{figure}[] 
\epsscale{0.9}
\plotone{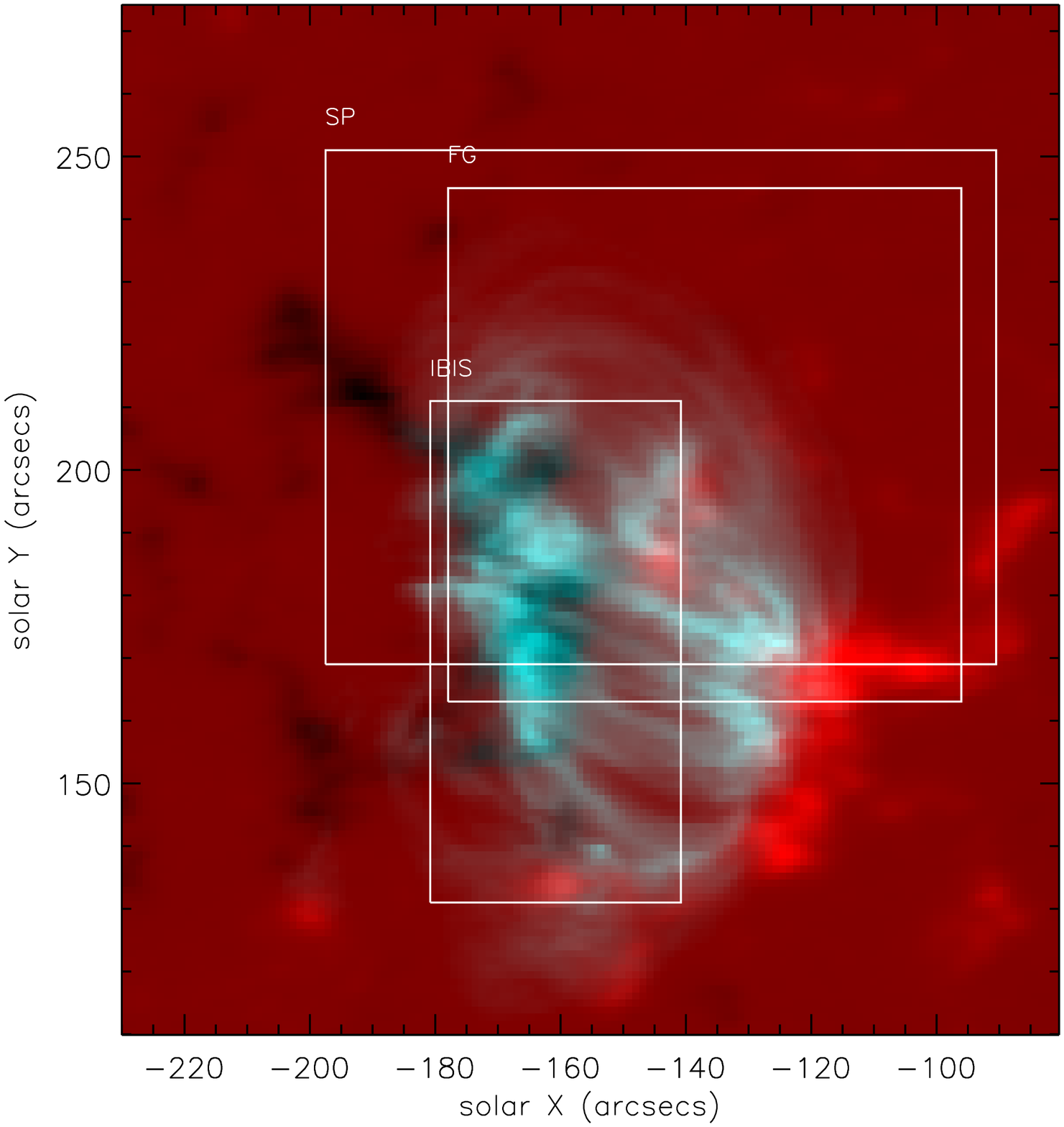}  
\caption{\label{fig:context}
A context image from \solis{} (red) and \trace{} (cyan) 
showing the areas observed 
around NOAA 10996 with the \hinode{} (spectropolarimeter- ``SP'', filtergram- ``FG'') and
with IBIS, before 15:06 UT on 20 May 2008.   The \solis{} scan began
at 15:11 UT, the \trace{} image shown was obtained at 15:12 UT.
The image coordinates refer to the \solis{} coordinate system.  
}
\end{figure}
}

\newcommand{\figtwo}{
\begin{figure}[] 
\epsscale{1.1}
\plotone{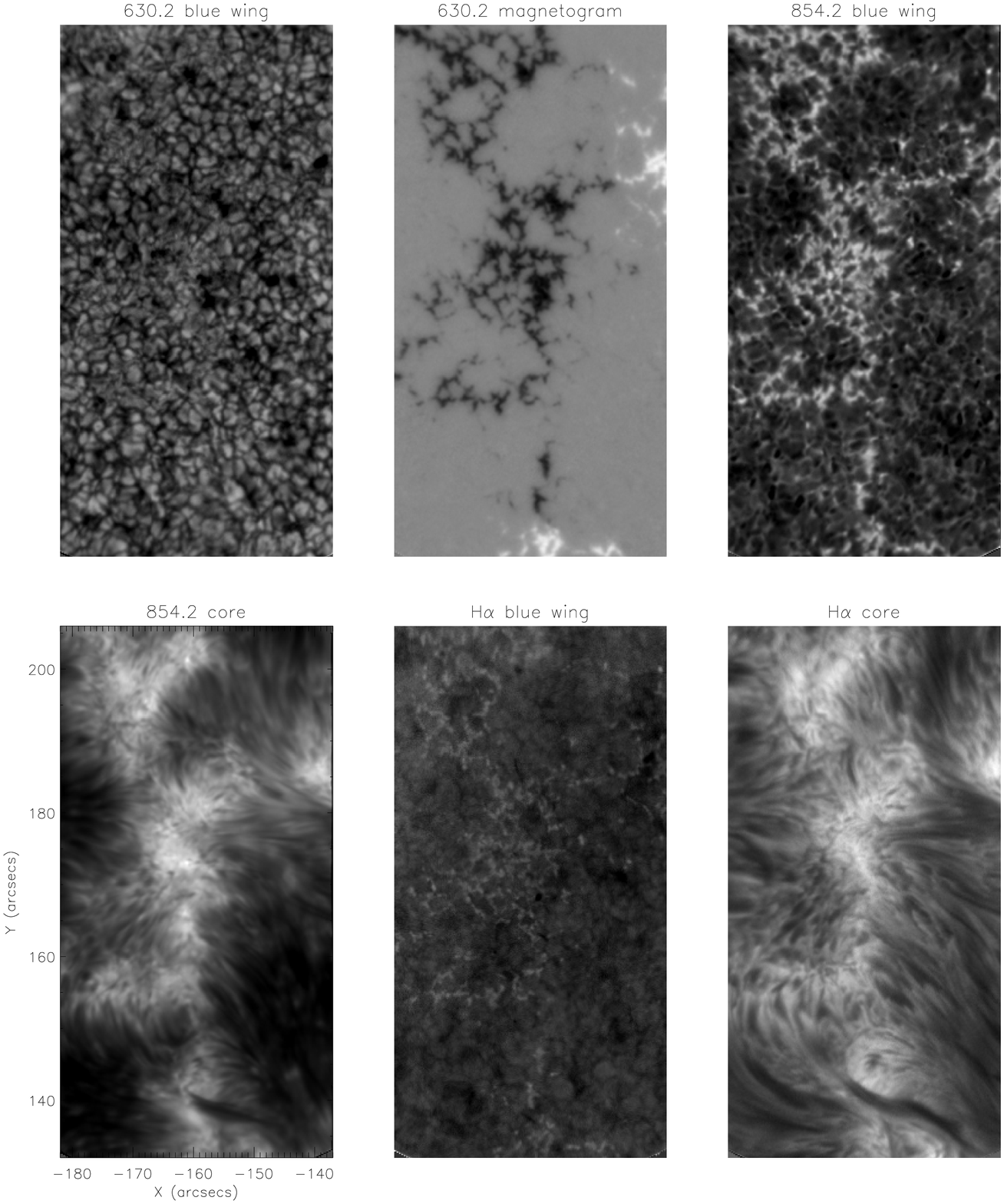}  
\caption{\label{fig:ibisone}
Typical \ibis{} data of NOAA 10996 observed on 
20 May 2008 from scan number 11. 
Almost the entire \ibis{} FOV is shown.  
The ``magnetogram'' is simply
$V$(-74 m\AA) minus $V$(+56 m\AA) for the \FeI{} line.
}
\end{figure}
}

\newcommand{\figthree}{
\begin{figure}[] 
\epsscale{1.0}
\plotone{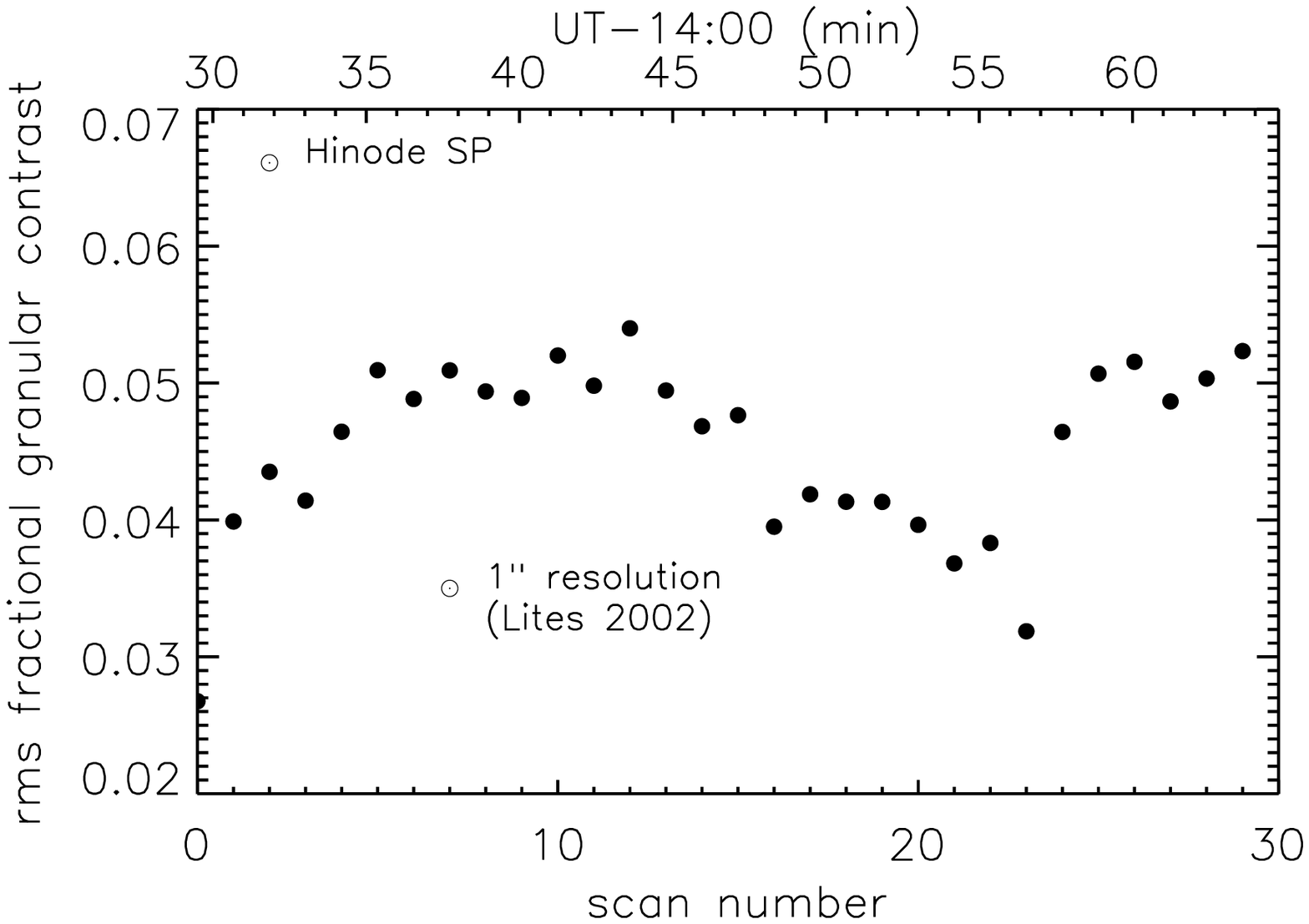}  
\caption{\label{fig:grancont}
RMS granular contrasts are shown for the quietest regions of the
\ibis{} and \hinode{} SP FOV, as measured in the continuum near 630.2
nm.  
Also shown is the contrast 
corresponding to an effective $1\arcsec$ resolution derived by
\protect\citet{Lites2002}. 
 }
\end{figure}
}

\newcommand{\figfour}{
\begin{figure}[] 
\epsscale{1.0}
\plotone{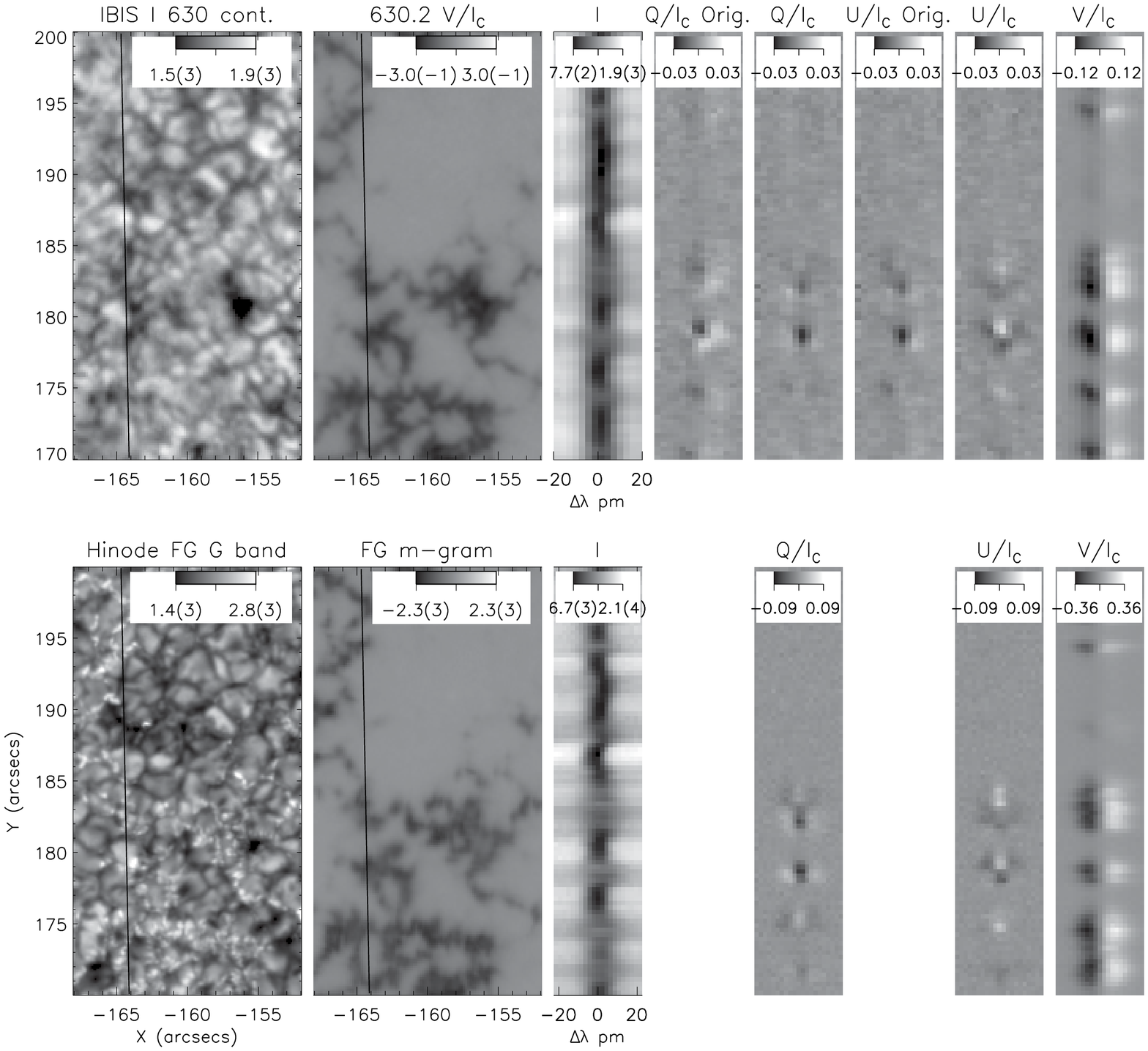}  
\caption{\label{fig:imspvsibis}
Typical \ibis{} and \hinode{} SP polarimetric
measurements of the 630.2 nm line.  The upper panels show \ibis{}
data, the lower show \hinode{} data.  The leftmost column shows intensity images, the next column 
magnetograms, from \ibis{} ($(V_{\lambda_0+6pm}-V_{\lambda_0-6pm})/I_C$, $\lambda_0=$ rest wavelength.) and the FG
instrument on \hinode{} (in Mx~cm$^{-2}$). 
The near-vertical
line shows the position of the \hinode{} slit at 14:36:52 UT, from which the six
rightmost images of Stokes profiles as a function of 
wavelength and position along the SP slit are taken.   $QU$ data
marked ``Orig.''  are before
subtraction of crosstalk and a final rotation in the polarization plane, the unflagged $QU$ data include these corrections.
\ibis{} profiles are extracted from 
the data closest in time and space to the SP measurements.  Color
scales are shown at the top of each image, where $x.y(z) \equiv
x.y\times10^z$.  
}
\end{figure}
}

\newcommand{\figfive}{
\begin{figure}[] 
\epsscale{1.}
\plotone{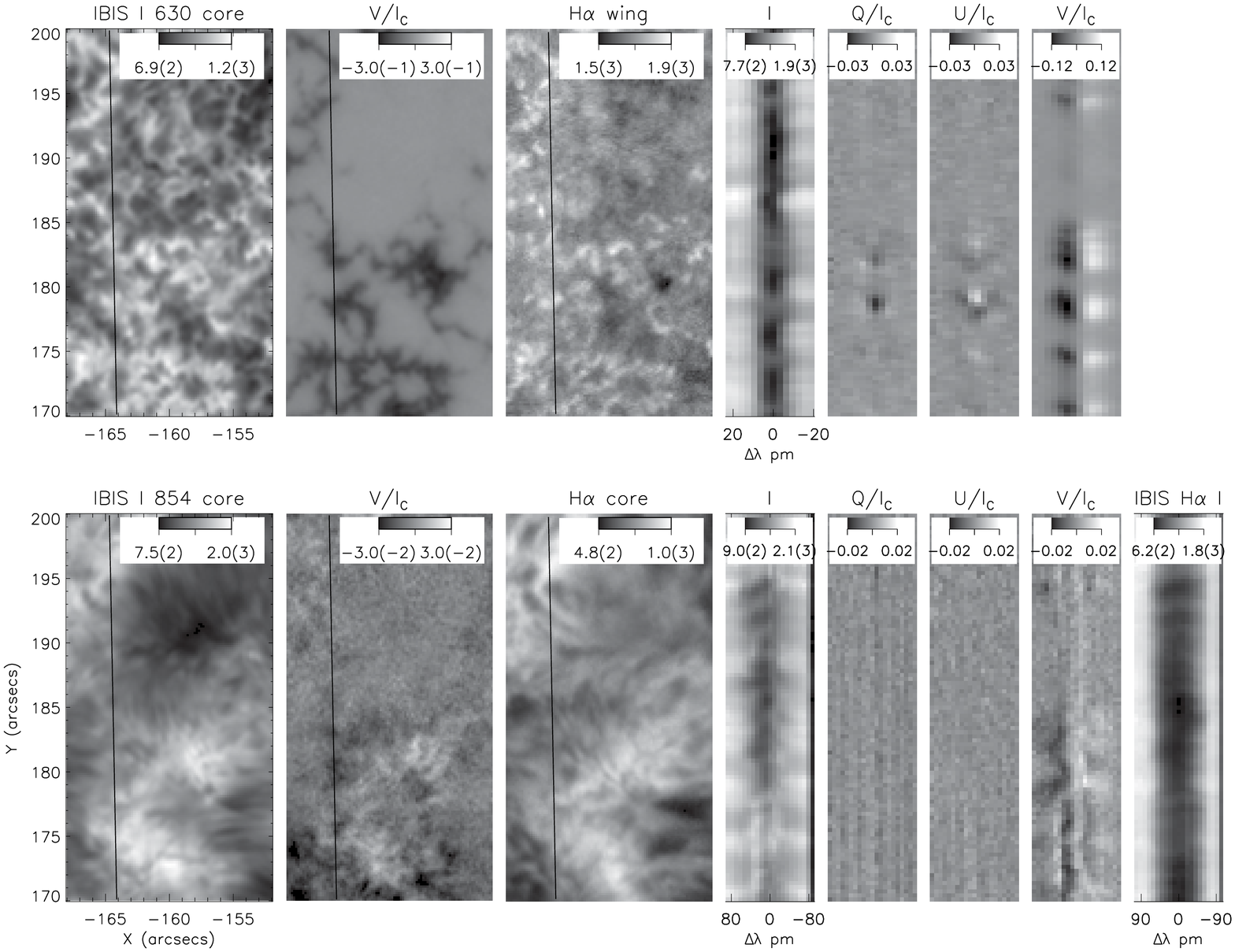}  
\caption{\label{fig:fe_vs_ca}
A figure similar to figure~\pref{fig:imspvsibis}, except that the
lower panels are \ibis{} data for the 854.2 nm line of Ca~II and
H$\alpha$, the $V/I_C$ frame for Ca~II shows 
$(V_{\lambda_0+17pm}-V_{\lambda_0-17pm})/I_C$.
 }
\end{figure}
}

\newcommand{\figsix}{
\begin{figure}[] 
\epsscale{1.}
\plotone{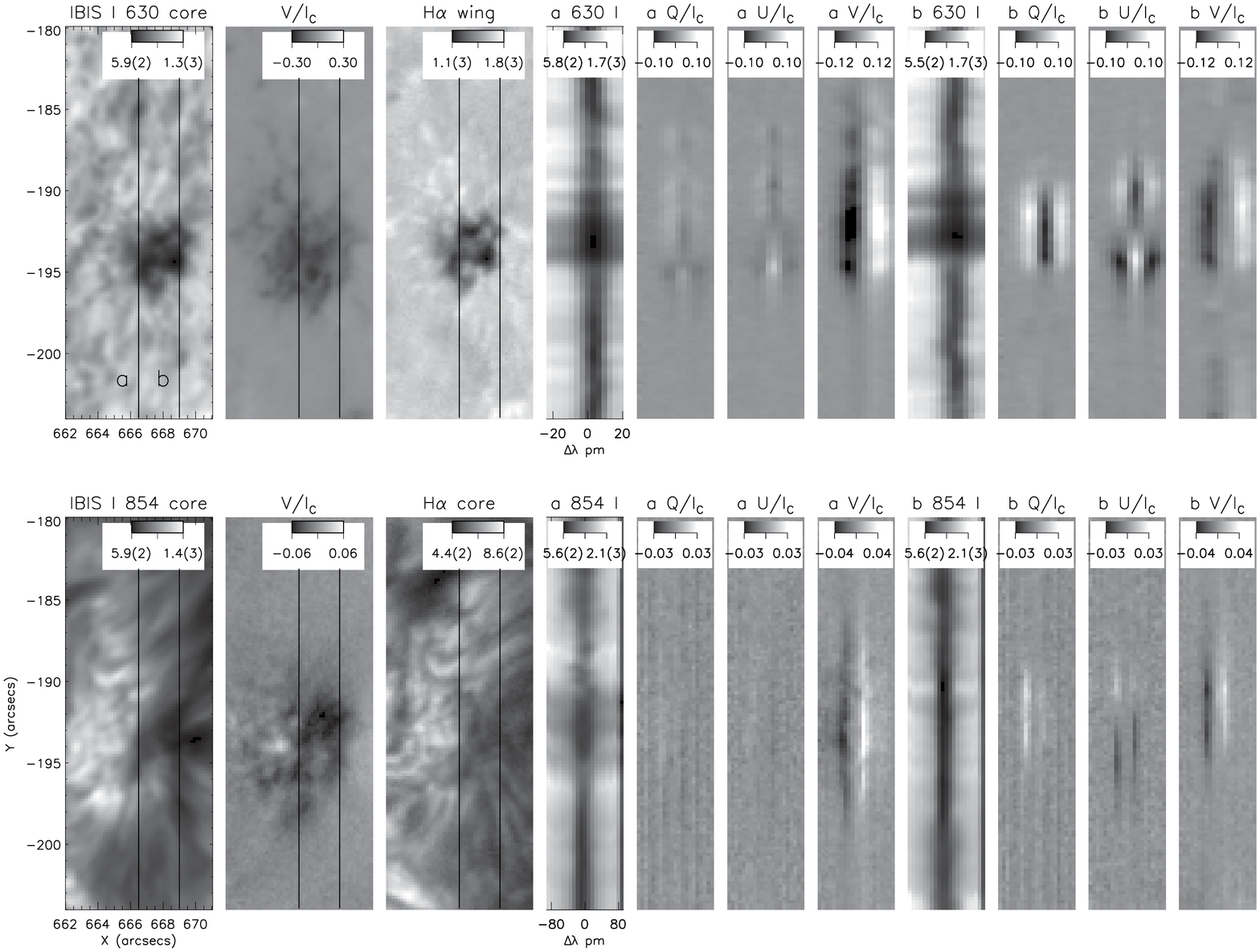}  
\caption{\label{fig:10994}
A figure similar to figure~\pref{fig:imspvsibis}, showing the 
pore region associated with  NOAA 10994 at 
S12.4, W46.1. 
 }
\end{figure}
}

\newcommand{\figseven}{
\begin{figure}[] 
\epsscale{1.}
\plotone{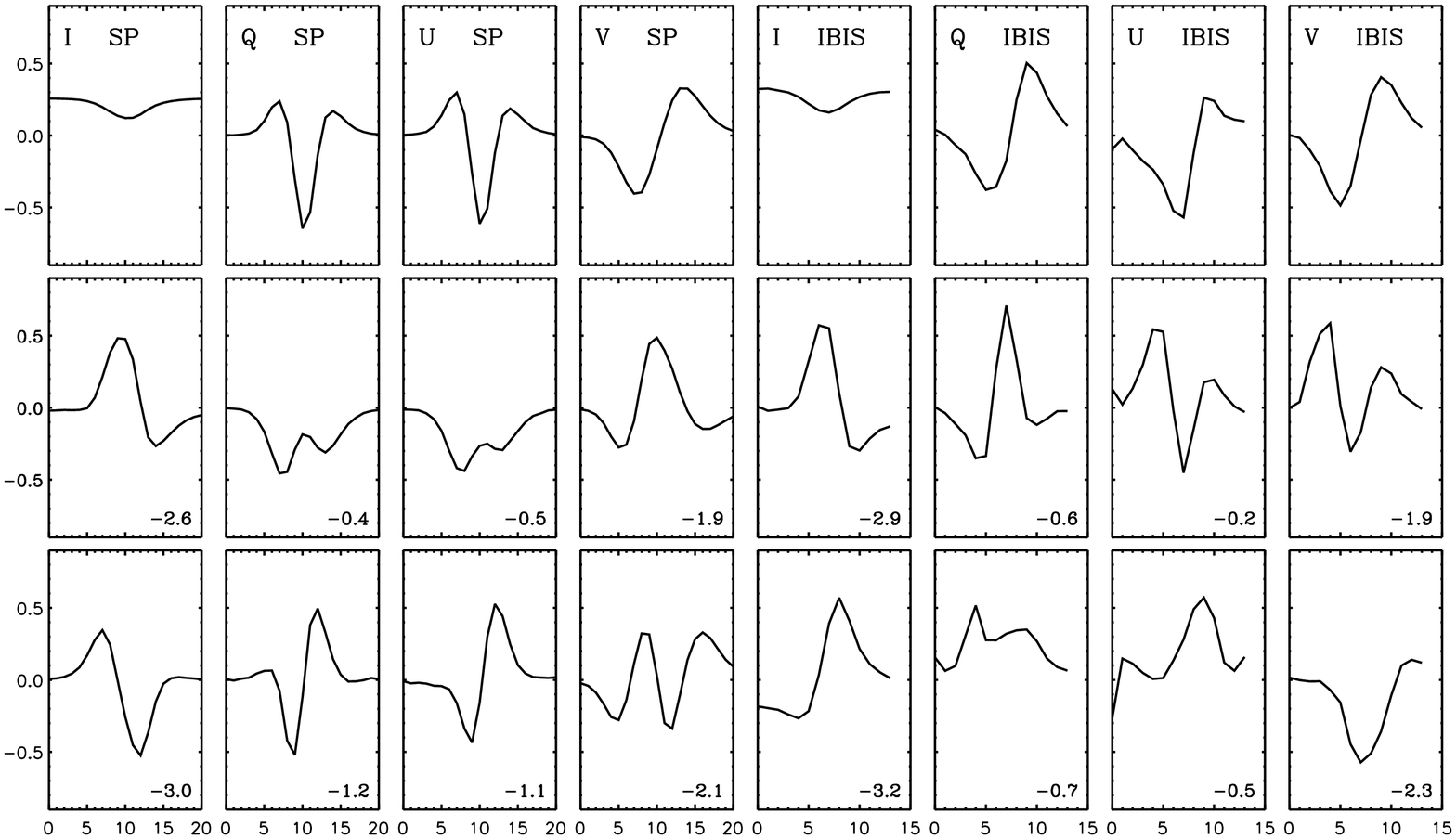}  
\caption{\label{fig:eigenibis}
The first three eigenvectors in the PCA expansion for each of the four
Stokes parameters of the Fe~I 630.2 nm line 
are shown, for the \ibis{} data from 14:43 UT on 20
May 2008, and the corresponding \hinode{} SP data.  
The first row is the eigenvector with the largest eigenvalue, the second and third show
the eigenvectors for the second and third largest eigenvalues.
The second and third rows lists  log$_{10}$ of the
modulus of the eigenvalue (the largest is set to 1). The columns are labelled with the instrument and Stokes parameter.
The abscissa is
wavelength index.
 }
\end{figure}
}

\newcommand{\figeight}{
\begin{figure}[] 
\epsscale{1.}
\plotone{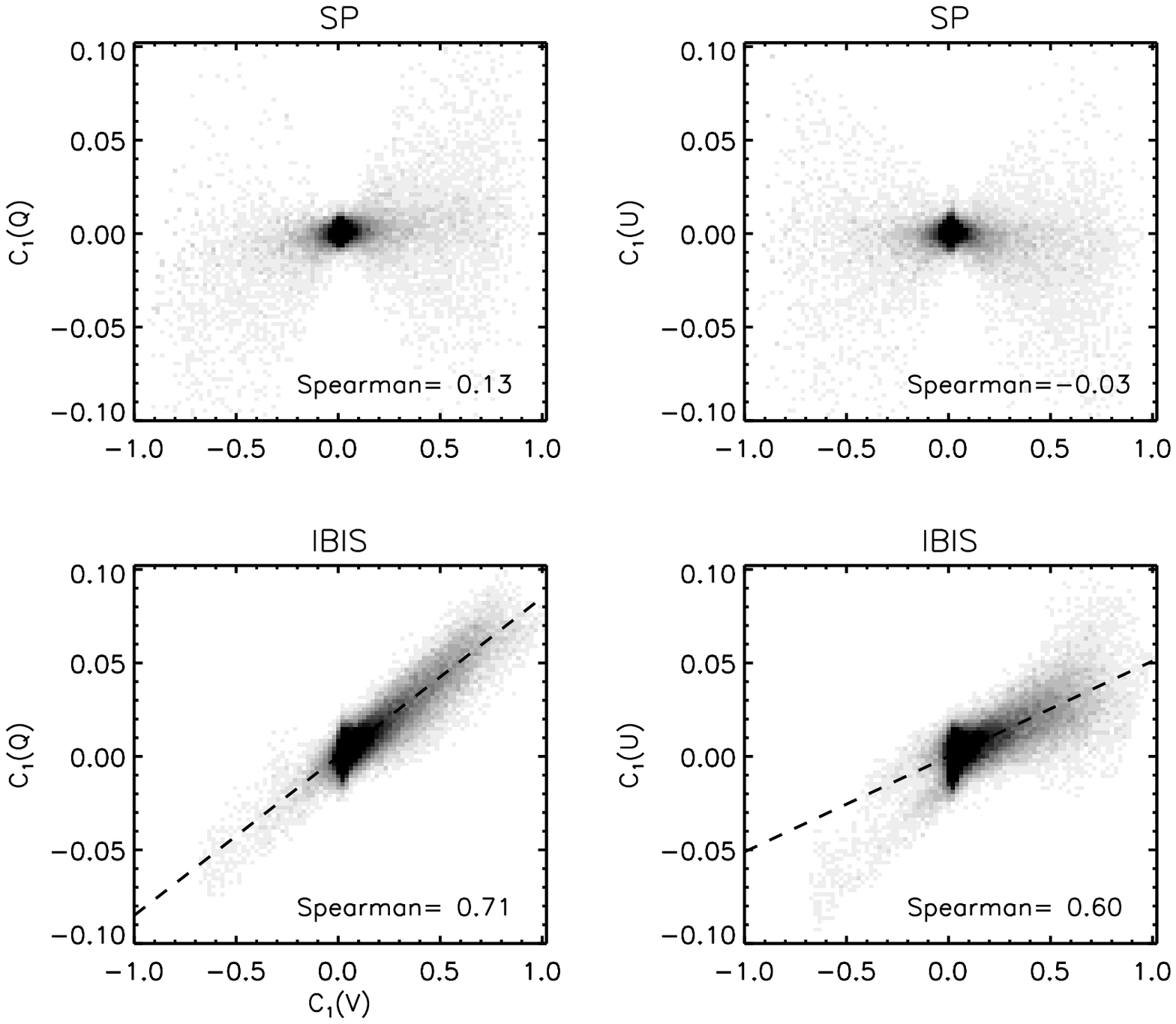}  
\caption{\label{fig:coeffs}
Scatter plot of the leading coefficients in the principal component
expansion for Stokes $QU$ and for $V$.   The dashed lines show the relations 
$c_1(Q)=0.085 c_1(V)$ and $c_1(U)=0.051 c_1(V)$ 
derived 
empirically using the symmetry properties discussed in the text.
 }
\end{figure}
}

\newcommand{\fignine}{
\begin{figure}[] 
\epsscale{1.}
\plotone{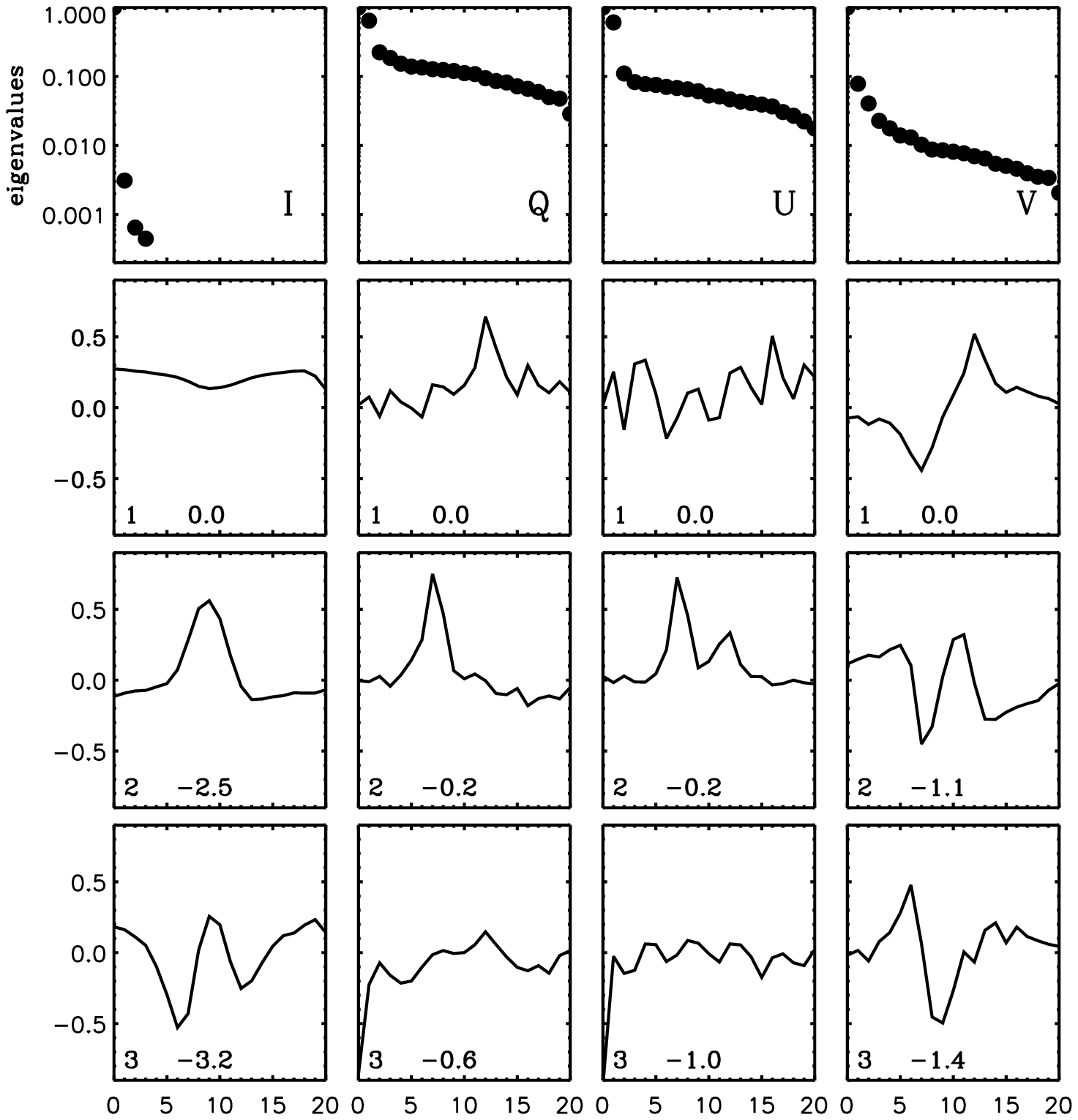}  
\caption{\label{fig:eigenibisca}
Eigenvalues and the first three eigenvectors 
from a PCA analysis of the 854.2nm 
line of Ca~II.  The data are constructed 
from the central $10\arcsec\times10\arcsec$ area of 
the pore, NOAA 10994, where significant linear polarization 
was measured.
}
\end{figure}
}

\newcommand{\figten}{
\begin{figure}[] 
\epsscale{1.}
\plotone{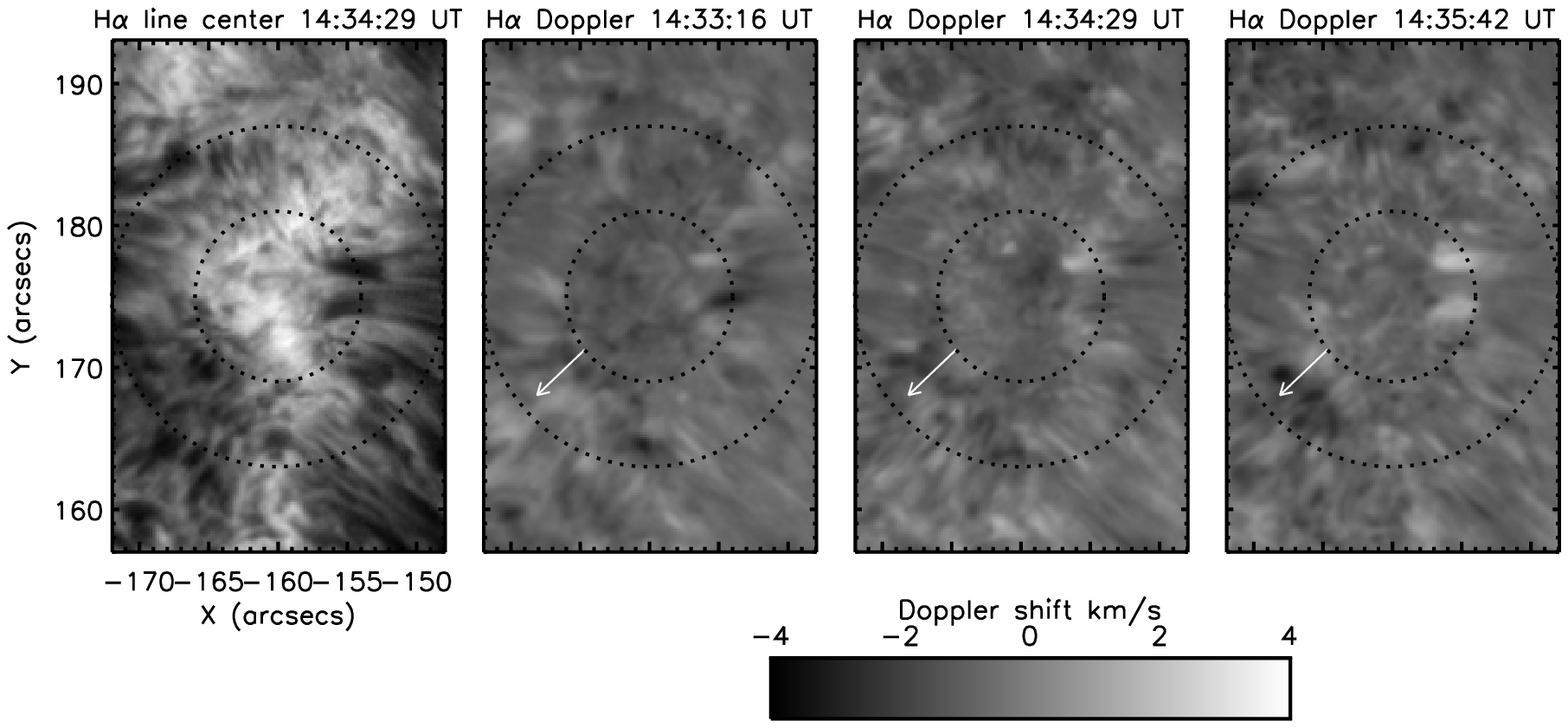}  
\caption{\label {fig:eiegenblobs}
H$\alpha$ data shown for NOAA 10996.  The leftmost image is a simple intensity image,
the others show the time sequence of three snapshots taken roughly 70s apart, of
Doppler shifts derived using the PCA technique.  Bright features are redshifted, dark features blue shifted. 
The dark blobs near 8 o'clock (see the arrow) propagate roughly outwards from the centers of the circles marking a bright concentration of magnetic flux.  
 These data are qualitatively compatible with 
magnetic field-directed flows along the magnetic fields originating from the flux concentration and expanding into the overlying corona. 
The velocity vectors can in principle yield the direction of the magnetic field. 
}
\end{figure}
}

\usepackage{graphicx}
\usepackage{colordvi}
\usepackage{color}

\newcommand{\philemail}{judge@ucar.edu}
\newcommand{\alfredemail}{dwijn@ucar.edu}
\newcommand{\aliemail}{ali@nso.edu}
\newcommand{\hanemail}{huitenbroek@nso.edu}
\newcommand{\kevinemail}{kreardon@arcetri.astro.it}
\newcommand{\giannaemail}{gcauzzi@arcetri.astro.it}

\shortauthors{Judge et al.}
\shorttitle{}

\slugcomment{}

\begin{document}

%
%

\title{Fabry-Perot versus slit spectropolarimetry of pores and active network. Analysis of IBIS
  and Hinode data}
\author{Philip G. Judge}
\affil{High Altitude Observatory,
       National Center for Atmospheric Research\footnote{The National %
       Center for Atmospheric Research is sponsored by the %
       National Science Foundation},
       P.O.~Box 3000, Boulder CO~80307-3000, USA; \philemail}
\author{Alexandra Tritschler, Han Uitenbroek} 
\affil{National Solar Observatory/Sacramento Peak\footnote{Operated by the %
       Association of Universities for Research in Astronomy, Inc. (AURA), %
       for the National Science Foundation}, P.O.~Box 62, 
       Sunspot, NM-88349, U.S.A.; \aliemail, \hanemail}
\author{Kevin Reardon, Gianna Cauzzi}
\affil{INAF -- Ossevatorio Astrofisico di Arcetri, I-50125 Firenze, Italy; 
       \kevinemail, \giannaemail}
\and
\author{Alfred de Wijn}
\affil{High Altitude Observatory,
       National Center for Atmospheric Research,
       P.O.~Box 3000, Boulder CO~80307-3000, USA; \alfredemail}


%
%

\begin{abstract}

  We discuss spectropolarimetric measurements of photospheric
  (\FeI~630.25\,nm) and chromospheric (\CaII~854.21\,nm) spectral
  lines in and around small magnetic flux concentrations, including a
  pore.  Our long-term goal is to diagnose properties of the
  magnetic field near the base of the corona.  We compare 
  ground-based two-dimensional spectropolarimetric measurements with 
  (almost) simultaneous space-based slit spectropolarimetry. We
  address the question of noise and cross talk in the measurements and
  attempt to determine the suitability of \CaII{} 
  measurements with imaging spectropolarimeters 
for the determination of chromospheric magnetic fields.
  The ground-based observations were obtained May 20, 2008, with the
  Interferometric Bidimensional Spectrometer (\ibis{}) in
  spectropolarimetric mode operated at the Dunn Solar Telescope (DST)
  at Sunspot, New Mexico. The space observations were obtained with
  the Spectro-Polarimeter of the Solar Optical Telescope (SOT) aboard
  the Japanese \hinode{} satellite.  The agreement between the
  near-simultaneous co-spatial \ibis{} and \hinode{} Stokes-$V$
  profiles at 630.25\,nm is excellent, with $V/I$ amplitudes
  compatible with to within $1\,\%$.  The \ibis{} $QU$ measurements
  are affected by residual crosstalk from $V$, arising from
  calibration inaccuracies, not from any inherent limitation of
  imaging spectroscopy.  We use a PCA analysis to quantify the
  detected cross talk.  $QU$ profiles with $V$ crosstalk subtracted
  are in good agreement with the \hinode{} measurements, but are
  noisier owing to fewer collected photons.  Chromospheric magnetic
  fields are notoriously difficult to constrain by polarization of
  \CaII{} lines alone.  However, we demonstrate that high cadence,
  high angular resolution monochromatic images of fibrils in \CaII{}
  and H$\alpha$, seen clearly in \ibis{} observations, can be used to
  improve the magnetic field constraints, under conditions of high
  electrical conductivity.
Such work is possible only with time series datasets from two-dimensional spectroscopic instruments such as \ibis{}, under conditions of good seeing.  
\end{abstract}


%
%

\section{Introduction}\label{sec:introduction}

This is first of several planned papers attempting to derive 
chromospheric magnetic structure using imaging spectropolarimetry, a new tool
with considerable potential.  Our motivation is to constrain the 
magnetic free energy in the solar atmosphere, the cause of many of
the Sun's most interesting observable phenomena, yet  measurements of
this free energy are notoriously difficult to obtain.  Line-of-sight
(LOS) components of photospheric magnetic fields have been measured
using circularly polarized light routinely for over 50 years. Such measurements set no constraints on the free magnetic energy.  It was not until credible measurements of the full polarization vector became available  \citep{Baur+House+Hull1980} that the {\em free} component of the magnetic energy became something amenable to observation.

But several difficulties arise.  The magnetic virial theorem relates the total magnetic free energy in a volume overlying a surface to the vector field measured at that surface, provided the surface is in a force-free state \citep[e.g.][]{Low+Lou1990}.  Most vector field measurements are made in photospheric lines where, outside of sunspot umbrae, the fields are far from force-free.  Thus one must either extrapolate the field into the overlying atmosphere, or make measurements higher in the atmosphere where the field is close to force-free (the ratio of gas to magnetic pressure, plasma $\beta$, $ \ll 1$).  The first option has been pursued by many researchers with mixed success
\citep{Schrijver+others2008,Wiegelmann+others2008,deRosa+others2009}, in essence because the problem is non-linear \citep{Sakurai1979}, 
is usually cast into force-free form incompatible with photospheric conditions, 
and is also
ill-posed (sensitive to boundary conditions, \citealp{Low+Lou1990}).

In this paper we begin pursuit of the second option using an imaging spectropolarimeter to observe chromospheric lines.  
Previous polarimetric studies of chromospheric lines have been 
successful in recovering magnetic fields \citep[e.g.][]{Metcalf+others1995,Solanki+others2003,Socas-Navarro2005a,Harvey2006,Centeno+others2006,Pietarila+others2007}, but are based on slit spectroscopy.
As we will show, imaging spectropolarimetry is especially suited to studying the chromosphere because it has the spatial coverage and high temporal cadence needed to follow dynamic fibril motions which mostly dominate the upper chromosphere \citep[e.g.][]{Judge2006}.  Such work is difficult with conventional slit spectrographs.
Outside of very quiet regions of the Sun, the plasma $\beta$ is certainly $ <1$ 
in the upper chromosphere,
as can be seen by inspection of H$\alpha$ spectroheliograms 
\citep[e.g.][]{Kiepenheuer1953,Athay1976} dominated by
fibril structures resulting from the entrainment of plasma by strong magnetic fields.  
By using measurements of the chromospheric vector field 
\begin{itemize}
  \item{}  the magnetic free energy in the overlying corona follows directly from the virial theorem;
 \item{}  extrapolations into the overlying atmosphere can be made using a boundary condition which is itself force-free, unlike the photospheric case;
 \item{} the change in regime from a forced
  (photospheric) to force-free state (upper chromosphere and corona) can be probed.
\end{itemize}
The third point is of more than academic interest. Recent work by
\citet{Leake+Arber2006,Arber+Haynes+Leake2007} has shown
 that ion-neutral collisions damp 
components of the current density ${\bf j}=$~curl~{\bf B} which are perpendicular to the magnetic field {\bf B}. A significant component of the free energy is thus dissipated in the chromosphere before it reaches the corona.

Below, we present 
vector polarimetry of the photosphere (\FeI\, 630.25\,nm) and chromosphere (\CaII\,854.21\,nm) using the imaging spectropolarimeter \ibis{}
\citep{Cavallini2006,Reardon+Cavallini2008}. 
However, while proven for earlier instruments \citep[e.g., the Imaging
Vector Magnetograph IVM, ][]{Mickey+others1996},
the technique needs to be proven for \ibis{}, 
and so 
we combine the \ibis{} data with (nearly) simultaneous 
measurements of the same \FeI{} line from the slit 
spectropolarimeter aboard the \hinode{} satellite, 
to assess the spectropolarimetric capabilities of \ibis{}.
Later work will present a physical analysis of these data together with H$\alpha$, 
EUV, X-ray and G band data obtained simultaneously.

%
%

\section{Observations}\label{sec:observations}

Joint observations with the Dunn Solar Telescope (DST), Hinode, and TRACE were obtained from 13 to 21 May 2008. Only data from 20 May 
2008 were of sufficient quality to merit further study and are reported here.
We
targeted the strongest magnetic features on the solar disk, 
but only small pores were observed. Two
regions were observed: the first (NOAA 10996), centered near solar
latitude and longitude N8.7, E10.2, was also observed by
\hinode{}. The second (NOAA 10994, S12.4, W46.1) was observed only at the DST.

Table~\pref{tab:obslog} summarizes the observations. The heliographic
pointings listed between instruments were determined from a \solis{}
full disk magnetogram. The \solis{} scan was obtained between 15:11:27
and 15:23:23\,UT, the region shown in Figure~\pref{fig:context}  being observed by
SOLIS near
15:16:15\,UT over a period of 1 minute.  The figure shows the
context of the NOAA 10996 \ibis{} and \hinode{} spectropolarimetric
measurements, in a \solis{} 630.2\,nm magnetogram and a \trace{}
19.5\,nm image. \trace{} 160\,nm  data obtained at 15:11\,UT were used
to determine the co-alignment with the \solis{} magnetogram, by
eye. The \solis{}-\trace{} alignment is accurate to $\pm 1\arcsec$.
The 19.5\,nm data shown have been corrected for the 160-19.5\,nm
offset using standard SolarSoft \trace{} software, and are from
15:12\,UT. Note that relative inter-instrument co-registration can be
made at sub-arcsecond scales, smaller than the uncertainties in
absolute heliographic coordinates. 

It can be seen that only the upper half
($\approx40\arcsec\times40\arcsec$) of the \ibis{} field of view
(FOV) overlapped the \hinode{} SP FOV.  It is also clear that \ibis{}
observed a region predominantly of negative polarity following some
50\arcsec{} behind a positive polarity region.  These small
concentrations of magnetic flux are connected by relatively short
coronal loops seen in the 19.5\,nm \trace{} data. There is some
19.5\,nm ``moss'' emission associated with the photospheric field
concentrations.  Moss is the bright footpoint emission at the base of
an overlying hotter corona  \citep{Fletcher+dePontieu1999, Berger+others1999}.

\subsection{ Spectropolarimetric Observations from \ibis}\label{subsec:dst}

The \ibis{} observations reported here were obtained between 14:29 and
15:19\,UT. \ibis{} is a Fabry-P\'{e}rot based filtergraph  in a
classical mount with a circular FOV of  diameter 80\,\arcsec,
rapidly tunable in wavelength 
\citep{Cavallini2006,Reardon+Cavallini2008}. \ibis{} 
was operated in dual-beam spectropolarimetric mode 
\citep{Tritschler+etal2009} in which 
two liquid crystal variable retarders (LCVRs) are 
placed in a collimated beam in front of the instrument, 
and a
polarizing beam-splitter is inserted in front of the detector. 
The FOV is then 
reduced to $\approx 40\arcsec\times80\arcsec$, because the 
beam splitter produces  a pair of simultaneous $I+ S_i$  and $I-S_i$ images
with $S_i=$ Stokes $Q$, $U$ or $V$\footnote{We will interchangeably use the notations
${\bf S}$, $(IQUV)$ and  $(S_0,S_1,S_2,S_3)$ as convenient henceforth, to describe the Stokes vector and its components.}.  At each wavelength
six modulation states were acquired sequentially, with the first beam 
acquired in the 
following order: $I+ Q$, $I- V$, $I- Q$, $I+ V$, $I- U$, and $I+ U$. 
This sampling order was chosen to maximize the efficiency of the LCVR modulation.
The \FeI, \CaII{} and H$\alpha$ lines were
sampled with  14, 21 and 22 non-equidistant wavelength points,
respectively.  
In H$\alpha$ only unpolarized measurements were
acquired.  Thus, a full sequence consists of $(14+21)\times 6 +
22=232$ exposures, requiring 70\,s in total.  For the
\FeI{} 630.2\,nm and \CaII{} 854.2\,nm lines the scanning was performed  by jumping
sequentially between the blue and red sides of the line.  To avoid
detector saturation the exposure time for each individual image was
set to 50\,ms, which limits the polarimetric sensitivity obtainable.
The instrument
performed electronic $2\times2$ binning prior to writing to  disk
which resulted in a detector image scale of 0.16\,arcsec\,pixel$^{-1}$.

The narrowband observations are supplemented with simultaneous wide
band data (WB, \@721\,nm) covering the same FOV with a detector
image scale of $\sim$0.08\,\arcsec\,pixel$^{-1}$. The WB data were used
for alignment purposes and the correction of the effects of
anisoplanatism  (to first order) during scanning via a de-stretch
algorithm using speckle reconstructed WB images as
reference images.  The speckle reconstructions were calculated from
bursts of 77 images each (burst durations of 23\,s)  using the
implementation by \citet{Woeger+vonderLuhe+Reardon2008}.  
G-band data covering a FOV of twice the width but the same
height as the WB data were also obtained  at a cadence of 0.2\,s on a
$1024\times1024$ detector with
$\approx$0.09\,\arcsec\,pixel$^{-1}$. These data are not discussed here. 
All observations were performed
in conjunction with the high-order adaptive optics (AO) system
\citep{Rimmele2004}. Seeing conditions were good but variable.  Full
calibration data sets including flats, darks, resolution targets  and
polarization calibration measurements were obtained before and after
science observations.

The \ibis{} $I\pm S_i$ data frames were reduced following procedures described by \citet{Cauzzi+others2008}, including dark subtraction, flat fielding, co-alignment with WB images, a blueshift correction needed because of the classical etalon mountings, de-stretching, and co-registration of all images in each sequence. The reduced $I\pm S_i$ frames were then combined and corrected for instrumental polarization to determine the solar Stokes vector {\bf S}.  The telescope calibration data used were from February 2007. A more recent calibration dataset was acquired in October 2008, but the results were not ready at the time of writing.  The measured {\bf S} is in a frame of reference defined by the elevation mirror of the telescope
\citep[see e.g.][]{Skumanich+others1997, Beck+others2005}. In the solar reference frame ($Q$ positive in the E-W direction on the Sun) a final rotation in the $Q-U$ plane is required.
However, prior to this rotation some care is needed because residual crosstalk from $V$ to $Q$ and $U$ is evident in the data. Thus an empirical correction to $QU$ using the $V$ data was applied, as described below, to try to derive the actual $QU$ entering the telescope, before applying the rotation.
The appendix describes the different origins of crosstalk in \ibis{} and the \hinode{} SP.  

For the first target, 30 \ibis{} scans of 70s each were completed, for the second, just 5 scans were obtained before clouds intervened.  Figure~\pref{fig:ibisone} shows typical \ibis{} filtergrams of NOAA 10996 taken in the blue wing of \FeI, \CaII{} and H$\alpha$, at the nominal line core position 
of \CaII{} and H$\alpha$, and a magnetogram constructed by subtracting $V$(+56 m\AA) from $V$(-74 m\AA) for the \FeI{} line.  The seeing was examined using a 10\arcsec{} square region centered near (-150\arcsec,120\arcsec), free of measurable magnetic influences.  Figure~\pref{fig:grancont} shows the rms intensity 
contrast measured with \ibis{}  and the \hinode{} SP in the continuum near 630 nm, as a fraction of the mean intensity.  
Also shown is the 630 nm continuum 
rms contrast corresponding to $1\arcsec$ resolution derived by \citet{Lites2002}.  The AO-corrected rms contrast was variable, and so was not dominated by stray light. The seeing was 
clearly worse during the first and 16th-24th \ibis{} scans.  The seeing-limited resolution of these \ibis{} observations, at best, corresponds to somewhere between $1\arcsec$ and the resolution of \hinode{} SP. (Note that ``resolution'' of \hinode{} here refers to the characteristic width of the point spread function, PSF, modified by the pixel size. In terms of granular contrast statistics, the smaller \hinode{} SP PSF corresponds to an effective resolution close to $0\farcs 315$.)

\subsection{Spectropolarimetric Observations from \hinode}

The Spectro-Polarimeter (SP, \citealp{Lites+Elmore+Streander2001}) aboard \hinode{}   \citep{Kosugi+others2007}  
obtained ``fast maps''  of 
the 630\,nm region.
In typical operation mode, 112 wavelength samples of 
the SP CCD detector are read, each pixel having a spectral width of 0.00215 nm and a width along
the slit of $0.16 \arcsec$ \citep{Centeno+Lites+deWijn2009}.  The 12-micron wide SP slit presents an effective 
width of $0.16 \arcsec$ to the solar image.  The fast map mode reduces telemetry volume 
via an effective 2x2 binning of the image produced by the SP mapping process: the image pixel
wells are binned in the direction along the slit during each read of the spectral/spatial 
CCD image, then successive integrations of one full rotation (1.6 sec) of the retarder polarization
modulator are summed onboard.  Successive integrations correspond to one step of
the image perpendicular to the CCD slit: an average step size of $0.149 \arcsec$.  
To further reduce the overall SP data rate for these coordinated observations as a
consequence of the failure in late 2007 of the Hinode onboard X-band telemetry 
system \citep{Shimizu2009}, one of the two CCD polarimetry images was not downlinked, 
and for the remaining data, we retained only the central half of the full $164 \arcsec$
length of the SP CCD along the slit.  The resulting SP fast maps for these observations
were then $335 \times256$ pixels, or approximately $99.6 \arcsec
\times 81.9 \arcsec$.
These maps required $\approx 21$ minutes to execute, and have an effective (square) pixel
aperture of about $0.3 \arcsec$.  With the Hinode rotating retarder modulator, demodulation 
was accomplished via onboard summing of images corresponding to 4 phases over a half-rotation
of the modulator. 
The Stokes vector was derived using the level 1 data product from the \hinode{} project which includes 
the 630.15\,nm and 630.25\,nm lines of \FeI.

While the SP observations are seeing-free, they do experience spacecraft jitter.  The amplitude of the jitter is remarkably small ($1\sigma< 0.01\arcsec$, \citealp{Shimizu+others2008}), which has two implications. First the images are far more stable than can be obtained from the ground. Second, the influence of jitter on the spectropolarimetry is small, when we understand jitter to have the same effect as ``seeing'' in the sense modeled by \citet{Lites1987,Judge+others2004}.  This issue is reviewed in the Appendix.

%
%

\section{Analysis}

\ibis{} obtains narrow-band 
2D images but must 
scan through multiple wavelengths to build up the spectra; the SP obtains spectra along the slit but must 
scan spatially perpendicular to the slit, to build up maps of the solar surface.  
The instruments also differ in the way the {\em polarimetry} is performed.  
\ibis{} is a ``Stokes definition polarimeter'', i.e.  during the
\ibis{} integrations only $I+S_i$ for $S_i=QU$ or $V$ is acquired.
Other than inevitable $I\rightarrow S_i$ crosstalk, which is mostly
removed by the dual beam, there is no other ``seeing induced
crosstalk''.  The combined modulation/demodulation matrices ($\tilde
H'_{ri}(\nu)$ of Lites 1987) are diagonal.  Yet such a polarimeter is
still susceptible to cross-talk 
because the telescope mixes the four polarization states prior to
entering the polarimeter in a fashion which may be imperfectly
calibrated, and the modulation/demodulation is not perfect. Crosstalk
induced by calibration errors varies far more slowly than seeing, and could be ``calibrated out'' if accurate calibration data were available.  Difficulties arise because the calibration matrices are imprecisely known.  Below, we will apply a simple correction for the $V\rightarrow QU$ crosstalk by simply requiring the average
$QU$ profiles to be symmetric around line center. The SP
data show this to be a good assumption.

Every SP exposure is a linear combination of $I$ and at least two other Stokes parameters.  Such measurements have significant off-diagonal $\tilde H'_{ri}(\nu)$ terms.  The SP is therefore subject to the systematic errors induced by ``seeing-induced crosstalk'' (again, for the SP, ``image motion'' means jitter).  But the residual image motion is very small: the factor $\beta_i$ in Lites' (1987) formula (15) is far less than it would have been if observing with the SP through the atmosphere.  It is in this sense that we can use the SP data as fiducial values against which we compare the data from \ibis{}.

\subsection{Polarization of \FeI{} 630.2 nm}
\label{subsec:fe1}
 
In Figure~\pref{fig:imspvsibis}, direct comparisons of Stokes parameters between \ibis{} and {\hinode} SP are shown. These data are typical of the entire dataset.  The intensity images and magnetograms show the context of the Stokes parameters shown in the rightmost panels, which for the \ibis{} data shown were were extracted from the near-vertical line representing the position of the SP slit.  First consider the Stokes $IV$ profiles.  The 
spatial variation of Stokes $I,V$ profiles along the SP  slit position
is remarkably similar
in the two instruments.  The higher angular and spectral resolution of the \hinode{} observations is evident.  Closer scrutiny reveals that Stokes $V$ to continuum intensity ratios, $V/I_c$, measured by \hinode{} are larger than those from \ibis{}, by a factor of typically 2.4.   
These differences can arise primarily 
because the magnetic field is spatially intermittent and not fully resolved.  Consider $V/I_c$ which 
measures the net LOS flux per unit area in each resolution element, in the weak field limit of the Zeeman effect \citep[e.g.][]{Lites2000}. This approximation not wildly incorrect for these data.  Consider just one magnetic element of area $a$ and LOS field strength $B_{los}$.  The magnetic field generates an intrinsic $V_{\rm actual}$.  But when measured by an instrument $i$ integrating over area ${ A_i} \ge a$, the Stokes $V$ is diluted by the (instrument-dependent) ``filling factor'' $f_i={ a}/{ A_i} \le 1$.  Then for instrument $i$, $V/I_c$ is 
\begin{equation} \label{eqn:stokesV}
V_i/I_c \propto f_i B_{los},
\end{equation}
\noindent Thus the lower the resolution (larger $A_i$), the smaller the $V/I_c$ signal, when the magnetic field is not resolved.  The presence of unpolarized 
stray light can also reduce $V/I_c$, but this effect is clearly smaller judging by the
measured granular contrast variations.
Similar arguments apply to $QU$ data. 
The data then suggest that the \ibis{} $V$ profiles sample areas $\lta
2.4$ times those sampled by the \hinode{} SP instrument,  thus the
effective resolution of these \ibis{} observations is $\lta1.5\times$
worse than that of \hinode{}. Taking, as noted above, the latter to be
$0\farcs3$ (and not twice this which is the Nyquist limit), we get
$0\farcs49$ for \ibis{}.  This number is compatible with the
seeing-limited resolution derived independently from
Figure~\pref{fig:grancont}.  Furthermore, the Stokes $V/I_c$
integrated over unipolar areas of several Mm$^2$ observed by \ibis{}
and \hinode{} SP give the same LOS magnetic {\em flux} (not flux
density) to within $1\%$, validating the above analysis, no unresolved flux of opposite sign being significant.
Below, we will show that the \ibis{} and \hinode{} Stokes $V$ data, as reduced into principal components, have very similar leading order eigenvectors.

\ibis{} $QU$ profiles are shown twice in Figure~\pref{fig:imspvsibis}.  Those marked ``Orig.'' are profiles obtained in the telescope reference frame.  These data are clearly not symmetric about line center, and appear $V$-like throughout.  Rotation of these profiles alone to the solar reference frame only involves linear combinations of the $QU$ data which will yield similar profiles with large asymmetric, $V$-like components.  The entire dataset is similarly contaminated by $V$-like profiles.  Some extra retardance has not been accounted for in the telescope calibration which converts incoming $V$ to $QU$ before entering the polarimeter.  Requiring that $QU$ be symmetric about the (Doppler-shifted) line centers, we found that the sign and amount of $V\rightarrow QU$ crosstalk is not a strong function of time, position on the detector etc.. 
Thus this crosstalk can be accounted for by a fixed Mueller matrix, of
which the important components lead to the corrections: 
$$Q=Q(Orig)-(0.085 \pm 0.004)V, \ {\rm and} $$
$$U=U(Orig)-(0.051\pm 0.003)V,$$ 
using $1\sigma$ statistical uncertainties.
Figure~\pref{fig:imspvsibis} shows these corrected $QU$ profiles also
rotated to the solar reference frame.  The agreement with \hinode{} SP
data for $QU$ is now remarkable.  These corrections are just the largest
terms arising from extra retardance $6^\circ$ oriented at about
$30^\circ$ to the \ibis{} reference direction, in which $QUV$ become
mixed via the Mueller algebra
\citep[e.g.][]{Seagraves+Elmore1994}. This retardance appears to arise
from a thin film of oil noted on the exit port of the telescope.  The
corrections above were applied to the lines of both Fe and Ca. 

The \hinode{} SP data for $QU$ in the magnetic network and pore are 
significantly above the designed sensitivity limit of the instrument.
They have essentially the symmetric profiles expected from the Zeeman
effect.
The heritage of \hinode{}'s SP is in the Advanced Stokes Polarimeter
\citep{Elmore+others1992,Skumanich+Lites+Pillet1994} and Diffraction
Limited Polarimeter \citep{Sankarasubramanian2006}.  Such profiles are
clearly of solar origin \citep[e.g.][]{Lites2000} and free of large systematic errors
introduced, for example, by the seeing-induced $V\rightarrow QU$
\citep{Lites1987,Judge+others2004}.

\subsection{Polarization of \CaII{} 854.2 nm}

Figure~\pref{fig:fe_vs_ca} shows a comparison of \FeI{} and \CaII{}
profiles for NOAA 10996, plotted as in figure~\pref{fig:imspvsibis},  
and from the same pixels.
At the peak of the $V$ profile near $y=179\arcsec$
the signal-to noise ratio of $V$ is $\sim$7 for the \CaII{} data
shown.  No significant $QU$ signal was detected in the \CaII{} data.
These data differ qualitatively from those of Fe~I, as found from the
analysis of Fe~I 849.7 and 853.8 nm lines observed with \spinor{} by 
\citet{Pietarila+others2007}.  Beyond 0.04 nm from line center,
i.e. at wavelengths where at least the intensity is formed in the
upper photosphere \citep{Cauzzi+others2008}, 
the \CaII{} $V$ profiles are similar to those of \FeI{} in their signs and spatial distribution along the slit.  Thus the nominal calibration of \ibis{} leads to credible Stokes $V$ signals in the parts of the \CaII{} line profiles whose intensities originate in the upper photosphere.  Hence, the (stronger) Stokes $V$ signals in the cores are also credible signals, formed in the chromosphere.  Within 0.04 nm of the line core, the profiles are spatially much more diffuse, reflecting an expansion of magnetic flux with height, but the core profiles themselves are more complex than their photospheric counterparts.

The ``magnetograms'' in the second column are constructed by
subtracting the blue lobes of each line's $V$ profile from the red
lobes, taking no account of Doppler shifts, at $\pm 0.0064$ and
$\pm0.017$ nm from line center for \FeI{} and \CaII{} respectively.
$V$ profiles show that \CaII{} ``magnetograms'' are a mix of Doppler, thermal and
magnetic signals which cannot be interpreted straightforwardly in
terms of LOS components of the chromospheric magnetic field.  Nevertheless
chromospheric magnetic fields are the origin of Stokes $V$ in \CaII{}
in this region, but there are also large influences from NLTE
radiative transfer and chromospheric dynamics
\citep{Pietarila+others2007}.  In particular, the \CaII{} line {\em
  intensity} forms over many scale heights \citep[e.g., see figure 5
of][]{Cauzzi+others2008}. Hence the inner wings form in the upper
photosphere where the magnetic field is relatively strong (compare the
\FeI{} line core data with the \CaII{} line wing data in
Figure~\pref{fig:ibisone}). But the core (within roughly $\pm 0.02$ nm
of line center) forms some 1Mm ( 6-7 pressure scale heights) higher,
in the middle to upper chromosphere.  The core intensity data 
\citep[see also figure 4 of][] {Cauzzi+others2008}
are dominated by magnetically-dominated fibril structures, over
concentrations of photospheric magnetic flux.  The question of the
formation heights of Stokes $V$ for the \CaII{} 854.2 nm line in such
regions is complex, and will be addressed in later work.  For now we
simply note that contributions to $V$ can arise from
the bulk of the chromosphere, spanning several pressure scale heights
and including the plasma $\beta=1$ surface.   Also the $V$
profiles below $y=178$ have two peaks and are correlated with strong
line core emission, but that from $y=179$ to $184$ the $V$ profile is
simpler and the cores are dark.  Evidently the relationship between
bright \CaII{} emission and magnetic field is not linear in our
data.  Similar results have been reported elsewhere
\citep[e.g.][]{Socas-Navarro+others2006}.

Calibrated data for NOAA 10994 are shown in Figure~\pref{fig:10994},
including profiles extracted from two columns (``a'' and ``b'').  The
local vertical of this region is at $47^\circ$ to the LOS, thus strong
vertical fields are seen partially as $QU$ signals from the transverse
Zeeman effect, as well as $V$.  Credible $QU$ profiles are observed in
the \CaII{} 854.2 nm line as well as the \FeI{} 630.2\,nm line.  These
show perhaps some residual $V$ to $QU$ crosstalk. This is not
surprising given the difference in wavelength between the Fe~I and
Ca~II lines and the unknown optical properties of the oil film.

\subsection{Principal Components}

Principal Component Analysis (PCA), a pattern/shape recognition
method, is useful in application to solar Stokes profiles because 
principal components (PCs) are often directly related to underlying
atmospheric
parameters \citep{Skumanich+Lopez2002}, for example
to magnetic field
strength, direction, line of sight motions, and thermal properties.
Here we use it also to quantify crosstalk.  PCs
are the eigenvectors of the covariance matrix with components defined
as
\begin{equation} \label{eqn:pc}
  {\cal C}_{ij} = \langle  S_i S_j \rangle_{{\bf x},t},
\end{equation}
\noindent where $S_i$ is the Stokes vector component (one of 
$IQUV$) at wavelength $i$, and
the angle brackets imply an average over all spatial pixels {\bf x}
and/or time $t$.   Denoting $e_{i}$ as
the $i^{\rm th}$ eigenvalue of ${\cal C}$ and $v_{i,j}$ as 
the $j^{\rm th}$ component of the eigenvector belonging to $e_i$, any data point
$d_{k,j}$, $k=$ spatial and/or temporal index, is represented by
\begin{equation} \label{eqn:expansion}
d_{k,j} = \sum_{i=1}^{n_\lambda}  c_{k,i} v_{i,j}
\end{equation}
\noindent where $n_\lambda$ is the number of wavelengths, and 
\begin{equation} \label{eqn:ccoeff}
c_{k,i} = \sum_{j=1}^{n_\lambda} d_{k,j} v_{i,j}.
\end{equation}
\noindent 
Assume that
the eigenvalues/ vectors are ordered in decreasing magnitude.  
When the eigenvalues $e_{i}$ drop steeply with increasing $i$, then most properties of the data are described by the first few eigenvectors in the sum.  The data points across the line profiles are not completely independent of one another, and so the data can be represented by truncating the sum to values smaller than $n_\lambda$.  If however, the spectrum is shallow, there is much independence between data points and a larger number of vectors are needed.  
In our \ibis{} data, Stokes $IV$ have steep spectra, but $Q$ and $U$
are shallower.  This is because $QU$ are noisy.  Noise introduces
linear independence between the different wavelengths across the
lines,
 thereby flattening the eigenvalue spectrum.  

Figure~\pref{fig:eigenibis} shows the first three eigenvectors derived
for SP and the \ibis{} dataset obtained near 14:43\,UT, covering the
FOV shown in \pref{fig:imspvsibis}.  The \ibis{} $QU$ data shown are
calibrated but uncorrected for cross talk because our aim is to
quantify the cross talk here using PCA.  Consider first the $QUV$ data
from \hinode{} in the figure.  The PCs for \hinode{} $Q$ and $U$ are
almost symmetric around the line center and resemble the second
derivative of the intensity profile, the principal $V$ component being
antisymmetric.  This is as expected because the \hinode{} $V$ and $QU$
profiles arise from the first and second order Zeeman effect,
respectively. While the \ibis{} principal $V$ component is asymmetric,
so too are the PCs of $QU$.   PCA has thus revealed the previously identified 
crosstalk in $QU$ as the principal signal, but it is to be noted that the more symmetric second and third eigenvectors have eigenvalues just 
0.6 and 0.2 dex below the PC's eigenvalue.   There is therefore significant signal 
of the appropriate symmetry in these \ibis{} data: the
second PCs of \ibis{} $QU$ resemble those seen with SP.  

Using PCA, $V\rightarrow QU$ cross-talk can be quantified by projection of profiles
onto the leading order PCs.
Figure~\pref{fig:coeffs} shows scatter plots
of leading order coefficients $c_{k1}$ for 
$Q$ and $U$ with those for $V$, for typical observations $k$ from both 
instruments.   There is no significant correlation for the \hinode{} SP
data, but the \ibis{} coefficients are correlated, the figure lists
Spearman rank correlation coefficients.  The dashed lines show $c_1(Q)=0.085 c_1(V)$ and
$c_1(U)=0.051 c_1(V)$, the numerical coefficients being those of
section \pref{subsec:fe1}.
If the $QU$ data were pure cross-talk from an antisymmetric $V$, then all points would 
be distributed along the plotted lines.  
The actual distributions show significant scatter, 
data for $U$ having a broader, skewed distribution. This behavior,
arising from an unknown source of systematic error,  
highlights the limitation of the simple {\em post-facto} corrections
of section~\pref{subsec:fe1}.  
The corrected $U$ profiles are 
probably reliable to at best $\pm 50$\%, given the plotted scatter.
Nevertheless, the PCA analysis lends support for 
our empirical corrections.

PCA should also help disentangle the complex \CaII{} $QUV$ profiles,
in terms of identifying the physical parameters most directly related
to the Stokes profiles \citep{Skumanich+Lopez2002}.
Figure~\pref{fig:eigenibisca} shows the eigenvalue spectra and first
few eigenvectors of the PCA expansion, for a
$10\arcsec\times10\arcsec$ area centered on NOAA 10994 (see
Figure~\pref{fig:10994}), a region chosen because it has measurable
linear polarization in the \CaII{} 854.2 nm line.  The $QU$ profiles
are mostly dominated by noise (as shown by the shallow gradient in the
eigenvalue spectrum, and noisy first
eigenvector). But the other eigenvectors show that genuine $QU$
signals are present for this small pore.  Stokes $V$ contains 
real signal because the first eigenvector is predominantly of
asymmetric form, and the eigenvalue spectrum is steeper than $Q$ and
$U$.  This $V$ component has no obvious net amplitude or area
asymmetry, in contrast with \citet{Pietarila+others2007}, who found a
net red-asymmetry in the 854.2\,nm line for some active network.

\subsection{Fibrils and their motions constrain the magnetic field}

Figure~\pref{fig:fe_vs_ca} also shows core and wing intensities of the
H$\alpha$ line from \ibis{}.  The wing and core behavior is similar to
the 854.2 nm line of \CaII{}, reflecting conditions at deeper layers
of the photosphere and top of the chromosphere respectively,
consistent with the picture of line formation of \citet{Athay1976}.
The fibril structures in the core intensity images are similar to, not
identical with, those for the 854.2 nm \CaII{} line.  H$\alpha$ images
at a given wavelength setting of \ibis{} are remarkably structured.
The motions of ``blobs'' of emission or absorption seem to trace the
fibril- and hence magnetic field line- orientations.  Curiously, on
the red side of H$\alpha$, blobs appear to converge on the underlying
photospheric flux concentrations.  On the blue side, they diverge from
them.  To try to determine if these ``blobs'' correspond to real
material motion, we show in figure~\pref{fig:eiegenblobs} H$\alpha$
data from three snapshots in the time series.  The leftmost panel is
the line core intensity image of the middle snapshot.  The three other
panels show data using PCA.  The particular eigenvector which
corresponds to line of sight Doppler shift, i.e. having a profile
corresponding to the first derivative of the intensity profile with
respect to wavelength, was identified.  This component was then
projected onto the data themselves at each point and time and images
displayed, yielding Doppler maps. 
(These Doppler velocities are
similar to those derived from the first wavelength moment of the
profiles).
Darker (lighter) profiles correspond
to blue (red) shifted components.  Examining the region between the
two plotted circles, several dark blobs of material move away from the
circle centers with time.  The Doppler shifts of these blobs are
towards the observer $(v_{los} <0)$. (While less clear, movies of the
red-shifted material show components converging onto the circle
centers).  This behavior is compatible with simple flows of absorbing
material which diverge from or converge to a magnetic structure
originating from the photospheric flux concentration and expanding
outwards, perhaps into the corona.  Hence, using proper motions (determined by eye in
this case) and Doppler shifts, one can trace out the 3D vector
velocity field of the entrained fibril material.  In the particular
blobs shown, the vector velocity has components
$(v_x,v_y,v_z)\approx(-7,-7,-3)$ km~s$^{-1}$, i.e. the velocity is
inclined at $17^\circ$ to the plane of the sky, only $16^\circ$ from
the local solar horizontal plane.  

These measurements are more than a curiosity.
Given the strong collisional coupling between the neutral and ionized
components, and the high electrical conductivity, the 3D velocity
field traces out the direction of the vector magnetic field, which is
therefore also only $\approx 16^\circ$ to the horizontal plane.  Such
observations offer observers the important opportunity to augment
spectropolarimetric measurements of chromospheric vector magnetic
fields, which will always suffer from weak Zeeman-induced $QU$
signals.  Measurement of the LOS component of the magnetic field from
Stokes $V$, coupled with kinematic fluid velocities, determines the
vector magnetic field.  Chromospheric fibril observations in Ca~II or
H$\alpha$ are thus far richer than suggested by inspection of
Figure~\pref{fig:fe_vs_ca}. They will be discussed in the context of
constraining the chromospheric magnetic fields elsewhere.

%
%

\section{Discussion}\label{sec:discussion}

We have demonstrated the fidelity of a two-dimensional filtergraph
instrument, \ibis{} for accurate Stokes $V$ measurements at high
angular resolution, by verification through almost simultaneous
measurements from the Spectro\--Polarimeter on board the \hinode{}
satellite.  \ibis{} measurements of Stokes $QU$ profiles of the \FeI{}
630.2 nm line are subject, as expected, to systematic errors (cross
talk), which we have quantified both empirically and 
using Principal Component Analysis.
\ibis{} is a Stokes Definition Polarimeter so that just one
polarization state is measured during each camera integration.
Therefore, the cross talk among $QU$ and  $V$ 
is of a slowly varying character,
originating from errors in the telescope calibration data.  An
empirical correction yields $QU$ profiles in remarkable agreement with
the SP data, yielding credible $QU$ profiles both in the \FeI{}
630.2\,nm line, and, for a pore far from disk center at a viewing
angle of $47^\circ$, credible observations of Stokes $QU$ in the
chromospheric \CaII{} 854.2 nm line.

We believe this is the first time that a direct comparison of
spectropolarimetry has been made between slit and two-dimensional
filtergraph instruments using data of the same region obtained within
a few seconds of each other. (The Advanced Stokes Polarimeter had this
capability by use of a slit-jaw but the results were not very good -
Lites 2009, private communication).  Given the intrinsic difficulty of
making spectropolarimetric measurements from the ground, the excellent
agreement of Stokes profiles found with near-simultaneous observations
from \ibis{} and the \hinode{} SP provides strong support for the
credibility of measurements made with an \ibis{}-like instrument. We
note that the
magnetic features observed here are rather weak compared with strong
sunspots, and the \ibis{} camera is to be updated to permit a larger
duty cycle and increased signal to noise.

Our goal is to diagnose properties of the magnetic field near the base
of the corona using, in part, the \CaII{} chromospheric lines. At face value,
the \CaII{} data appear to be of limited use in that under many
conditions the noise is dominant in the $QU$ data and corrections must
be applied for inaccuracies in the telescope polarization calibration
data.  Other complications include difficulties concerning non-LTE
transfer of in the \CaII{} lines.  Yet \ibis{} has some significant
advantages over slit spectropolarimetric measurements.
 \ibis{} images can be corrected using image reconstruction
techniques.  This is important because an angular resolution of
$1\arcsec$ or better, covering fields of tens of seconds of arc 
with a spectral resolution of some tens of
m\AA{} is required to see clearly the fibril structure of the upper
chromosphere (see Figure~\pref{fig:ibisone}).  A cadence of 70
seconds is barely enough to track kinematic motion along the fibrils, 
$\lta 20$ seconds being highly desirable.  Such criteria may not be
achieved using slit instruments.
The observation of fibril kinematics, even in unpolarized light,
appears important for diagnosing chromospheric magnetic fields because
fibrils not only trace out magnetic lines of force, but the partially
ionized plasma is required to flow along these lines because of the 
large electrical conductivity.  From
our \ibis{} data for H$\alpha$, it appears possible to measure the
velocity vector of fibril ``blobs'' in the plane of the sky, and along
the LOS, yielding the direction of the magnetic field (to within an
ambiguity of 180$^\circ$). Coupled with the Stokes $V$ signals which
may in principle yield the LOS field strength, there is thus hope that
the vector magnetic field might be constrained from such observations.

In conclusion, this dataset provides, for the first time, credible
imaging spectropolarimetry from photosphere through the chromosphere
together with a useful time series of resolved fibril motions in the
upper chromosphere.  These data can be used to place observational
constraints on the vector magnetic field throughout the chromosphere.
In the case of NOAA 10996, we have found that the magnetic field is
highly inclined to the LOS, as in a traditional magnetic ``canopy''
\citep{Giovanelli1982}, and that the chromospheric heating rates are
not simply related to the Stokes $V$ profiles of \CaII{}.  We will
explore the possibilities further using the data described here in
later publications.  More generally, \ibis{} holds promise as a tool
for chromospheric vector polarimetry in the \CaII{} 854.2 nm line,
following the work on a larger sunspot by
\citet{Socas-Navarro2005a}.

%
%

\acknowledgements

PGJ gratefully acknowledges the help of F. Woeger in speckle reconstruction, and discussions with B. W. Lites whose comments helped to improve the manuscript.
Hinode is a Japanese mission developed and launched by ISAS/JAXA,
collaborating with NAOJ as a domestic partner, NASA and STFC (UK) as
international partners. Scientific operation of the Hinode mission is
conducted by the Hinode science team organized at ISAS/JAXA. This team
mainly consists of scientists from institutes in the partner
countries. Support for the post-launch operation is provided by JAXA
and NAOJ (Japan), STFC (U.K.), NASA (U.S.A.), ESA, and NSC (Norway).
This research has also made use of NASA's Astrophysics Data System
(ADS).  An anonymous referee helped us improve the presentation.

\appendix

\section{Origins of crosstalk}

Consider for simplicity the case of a perfect polarimeter.  Let {\bf M} be the Mueller matrix describing the known telescope calibration parameters.  The Stokes vector {\bf
  S} entering the polarimeter is
$
{\bf R} = {\bf M S}
$
where ${\bf S}$ is the (desired) solar Stokes vector.   {\bf R} 
is inferred from the modulation and analysis of instrument $i$.  Then the estimate of the solar Stokes vector from this instrument is 
$$
{\bf S}^i = {\bf M}^{-1} {\bf R}
$$
With (fictional) perfect knowledge of the Mueller matrix, the solar Stokes vector is 
$$
{\bf S} = {\bf M_0^{-1} R}
$$
We can write $ {\bf M} = {\bf M_0 + E}$ where the (unknown) elements of the error matrix {\bf E} are assumed small compared with ${\bf M}_0$- the telescope calibration matrix is close to the actual matrix.  Keeping first order terms only, 
$$
{\bf S} - {\bf S}^i ={\bf E M^{-1} } {\bf R} \equiv {\bf E S}^i,
$$
which is the error in the inferred solar Stokes vector resulting from an inaccurate telescope calibration.   Off-diagonal (unknown) terms in {\bf E} convert the inferred ${\bf S^i}$ values into other Stokes components.   For a stable instrument, {\bf E} can be considered 
constant or very slowly varying with time. 
This kind of error is present in all practical polarimeters.

\newcommand{\blurred}{T}

In the presence of atmospheric seeing, the Stokes vector entering the telescope is a rapidly varying distortion of ${\bf S}$ in which neighboring points on the solar surface are spatially smeared.  Let $\blurred_\alpha(x,y;t)$ represent the $\alpha$ component of the Stokes vector entering the telescope from apparent position $(x,y)$ on the sky at time $t$.  Then, including just the lowest order (tip/tilt) distortions at any time $t$, ${\bf \blurred}$ arises from a different place on the sky $(x',y')=(x+\delta x(t), y+ \delta y(t))$.  Expand 
$\blurred_\alpha(x,y;t)$ 
as a Taylor series in the solar Stokes component ${S}_\alpha$ on the sky:
\begin{eqnarray} \label{eqn:atimed} \nonumber
\blurred_\alpha(x,y;t) = S_\alpha(x+\delta x,y+\delta y;t) 
\approx
S_\alpha(x,y;t) +  \nabla S_\alpha \cdot {\bf s}(t)  \label{eqn:seeing}
\end{eqnarray}
to first order \citep{Judge+others2004}.     Here the $\nabla$ operator and vector 
${\bf s}(t)$ are vectors in the $(x,y)$ plane,   ${\bf s}(t) = (\delta x(t),\delta y(t))^T$ is the tip-tilt component of the 
seeing.  Let $P_\nu$ be the power spectrum of the seeing.
All measurements require a finite integration time $\tau$, during which the 
seeing ${\bf s}(t)$ varies.  At all frequencies $\nu$  where $P_\nu \tau \ge 1$, seeing 
detrimentally affects the measurements.    When integrated over time $\tau$, detectors record energy
\begin{equation}
  \label{eq:energy}
\int_0^\tau  \sum_{\alpha=0}^3  a_\alpha \blurred_\alpha(x,y;t)  \, dt=
\int_0^\tau 
\sum_{\alpha=0}^3  a_\alpha \left \{ S_\alpha(x,y;t)  +   \nabla S_\alpha \cdot {\bf s}(t) 
\right \}\, dt,
\end{equation}
where $a_0=1$ ($S_0=I$) and $a_{\alpha \ne 0}$ depends on the polarimeter's particular modulation and integration scheme.  With many realizations of the seeing, the 
equation becomes 
\begin{equation}  \label{eqn:averages}  
\sum_{\alpha=0}^3  a_\alpha \langle \blurred_\alpha \rangle = 
\sum_{\alpha=0}^3  a_\alpha  \langle S_\alpha \rangle, 
\end{equation}
because $\langle \nabla S_\alpha \cdot {\bf s}(t) \rangle$  averages to 
zero when ${\bf s}(t)$, arising from atmospheric turbulence, is statistically spatially symmetric.   
The $\langle S_\alpha \rangle$ components have variances arising from the seeing:
\begin{equation}
  \label{eq:variances}
\sigma^2_\alpha = \langle |  \nabla S_\alpha \cdot {\bf s}(t) |^2 \rangle
\end{equation}
which \citet{Lites1987} has shown can be evaluated from the seeing power spectrum 
using the assumption that the seeing is a random phenomenon.  
Equation~(\pref{eqn:averages}) is a linear system for the desired 
solar components $\langle S_\alpha \rangle$ in terms of the measurements
$\langle \blurred_\alpha \rangle$, subject to statistical variations described by 
$\sigma_\alpha$.

For a Stokes definition polarimeter like \ibis{}, $a_{\alpha=1,2,3}$ is non-zero only for one Stokes parameter, $\beta$ say.  Then the above equations relate $\langle S_0 \rangle +a_\beta \langle S_\beta \rangle $ only to $\langle \blurred_0 \rangle$ and $a_\beta \langle \blurred_\beta \rangle$.  The second beam yields simultaneous measurements of equation $\langle \blurred_0 \rangle -a_\beta \langle \blurred_\beta\rangle$, and this two equation system is solved for all components $S_\beta$ in terms of $\blurred_\alpha$ as usual, with uncertainties propagated via $\sigma_0$ and $\sigma_\beta$.  

But for other polarimeters, like the SP on \hinode, $a_\alpha$ is non-zero for at least two of $QUV$ during each measurement.  The above equation becomes a system of more than two equations. After eliminating the intensity from the equation using the second beam we see that Stokes parameter $S_\beta$ is also influenced though the terms $\sigma_{\alpha \ne \beta}$.  If the variances $\sigma_\alpha$ (e.g. Stokes $V$) are larger than the desired terms $S_\beta$ (e.g. Stokes $Q$), then the linear system yields estimates for $S_\beta$ which are dominated in any particular realization by terms of order $\sigma_\alpha$.  In words, crosstalk arises because, during the exposure for modulation state $S_0+S_\beta$, fluctuations in time occur in $S_{\beta \ne \alpha}$ resulting from the motion of the solar image with contains structure.  The great advantage of the SP is its remarkably small rms image motion, so that image-motion induced cross talk is practically negligible.

\tabone
\clearpage

\figone
\figtwo
\figthree
\figfour
\figfive
\figsix
\figseven
\figeight
\fignine
\figten
\end{document}